\documentclass[conference,a4]{IEEEtran}
\IEEEoverridecommandlockouts
\usepackage{cite}
\usepackage{array}
\usepackage{amsmath,amssymb,amsfonts}
\usepackage{algorithm, algpseudocode}
\usepackage{subcaption}
\usepackage{graphicx}
\usepackage{textcomp}
\usepackage{xcolor}
\def\BibTeX{{\rm B\kern-.05em{\sc i\kern-.025em b}\kern-.08em
    T\kern-.1667em\lower.7ex\hbox{E}\kern-.125emX}}
    
\usepackage{stfloats}
\usepackage{float}
\usepackage{verbatim}
\usepackage[short]{optidef}

\begin{document}

\title{Sleep or Transmit: Dual-Mode Energy‑Efficient Design for NOMA-Enabled Backscatter Networks}

\author{\IEEEauthorblockN{Hajar El Hassani\IEEEauthorrefmark{1},~Member, IEEE, and Mikael Gidlund\IEEEauthorrefmark{2},~Fellow Member, IEEE}

\\[0.1in]

\IEEEauthorblockA{\small
\IEEEauthorrefmark{1}ETIS, UMR 8051, CY Cergy Paris University, ENSEA, CNRS, Cergy, France\\
\IEEEauthorrefmark{2}
Department of Computer and Electrical Engineering (DET), Mid Sweden University, 85170 Sundsvall, Sweden\\ 
Email: hajar.el-hassani@ensea.fr, mikael.gidlund@miun.se}

\thanks{}}

\markboth{}%
{Shell \MakeLowercase{\textit{et al.}}: A Sample Article Using IEEEtran.cls for IEEE Journals}

\maketitle

\begin{abstract}
The rapid growth of Internet-of-Things (IoT)
devices demands communication systems that are both
spectrally efficient and energy frugal. Backscatter communication (BackCom) is an attractive low-power paradigm, but its spectral efficiency declines in dense deployments. This paper presents an uplink BackCom design that integrates non-orthogonal multiple access
(NOMA) and maximizes system energy efficiency
(EE). In a bistatic network where multiple backscatter
nodes (BNs) harvest RF energy and alternate between sleep and
active modes, we formulate a fractional program with
coupled time, power, and reflection variables and develop a Dinkelbach-based alternating optimization (AO) algorithm with closed-form updates. Analysis reveals two operating modes depending on power availability, circuit demands and propagation conditions. Simulations show the proposed design adapts the time allocation, achieving up to $8\%$ higher
EE than fixed-power and $68\%$ than no-sleep baselines, and delivering up to $127\%$ EE gains over orthogonal multiple access (OMA). These results establish
NOMA-enabled BackCom as a scalable, energy efficient
solution for large-scale IoT deployments.
\end{abstract}

\vspace{-0.1in}
\section{Introduction}
\IEEEPARstart{T}{he} explosive growth of the Internet of Things (IoT) is projected to exceed 24 billion connected devices by 2030 \cite{iot-2030}, creating a pressing need
for wireless communication technologies that are simultaneously
scalable and energy efficient. Conventional wireless networks, which rely on power-hungry transceivers and active RF generation, are ill-suited
to large populations of battery-free sensors operating
under stringent energy constraints.

Backscatter communication (BackCom) has emerged as a promising ultra-low-power paradigm in which a device modulates and reflects an incident RF carrier rather than generating its own signal \cite{backcom_surv1,backcom_surv2}. BackCom architectures include monostatic, bistatic, and ambient variants, which differ in the placement of the RF source and receiver and in their use of dedicated or ambient carriers \cite{iot-2030,BackCom-categories,BackCom-categ2}. By eliminating high-power amplifiers and analog-to-digital converters, BackCom
enables battery-free IoT nodes, but its spectral efficiency and scalability degrade when many devices
share the channel.

Non-orthogonal multiple access (NOMA) offers a natural complement to BackCom. By exploiting power-domain multiplexing and successive interference cancellation
(SIC), NOMA allows multiple backscatter nodes (BNs) to transmit concurrently on the same time–frequency resource, improving spectral and energy
efficiency (EE) compared with orthogonal schemes such as time-division multiple access (TDMA) or frequency-division multiple access (FDMA) \cite{Elhassani20,BackCom-NOMA,rate_NOMA_ding,survey_NOMA_backcom1}. Several studies have examined BackCom with either OMA or NOMA, investigating metrics such as throughput, bit-error rate, or downlink EE \cite{EE_BBCom_OMA, backcom_noma_uplink1,BER_NOMA_BBcom,rate_NOMA_ding,EE_NOMA_BackCom,rate_NOMA_BackCom,elhassani_globecom21,elhassani_globecom22,elhassani_TGCN}. However, most of these works optimize a single resource, such as the reflection coefficient or transmit
power, or assume fixed sleep–active patterns, leaving
open the joint design of energy harvesting, time allocation, and RF power control in uplink NOMA-enabled
BackCom networks.  

In this work, we propose an energy-efficient uplink
BackCom framework that integrates NOMA to support
multiple batteryless BNs. The network operates in a bistatic configuration where an RF source powers the BNs and serves as the carrier for backscatter modulation. We formulate a joint design that simultaneously optimizes the RF-source transmit power, the sleep/active time allocation, and each BN’s reflection coefficient to maximize system EE. A Dinkelbach-based alternating-optimization (AO) algorithm is developed to solve the resulting non-convex
fractional program and to provide closed-form updates. The key contributions are as follows:

\begin{itemize}
    \item We formulate a joint optimization of RF-source transmit power, time-split between sleep and active phases, and BNs reflection coefficients for maximizing EE in NOMA-enabled BackCom.
    \item We develop a Dinkelbach-based AO algorithm with closed-form updates and provable convergence and complexity analysis. We identify two fundamental operating modes; harvest-on-transmit (HoT) and harvest-then-transmit (HtT), and their dependence on RF source power and BNs circuit demands. We also highlight the impact of these modes on the system latency.
    \item We demonstrate, via extensive simulations, the  effect of power availability, circuit demands and propagation conditions on the performance. We also highlight the substantial EE gains over OMA and baseline schemes, particularly in low to moderate power regimes.
\end{itemize}

The remainder of the paper is organized as follows.
Section II describes the system model, Section III
presents the optimization framework, Section IV reports
the simulation results, and Section V concludes
the paper.

\section{System model}

\begin{figure}[t!]
    \centering
    \includegraphics[width=\linewidth]{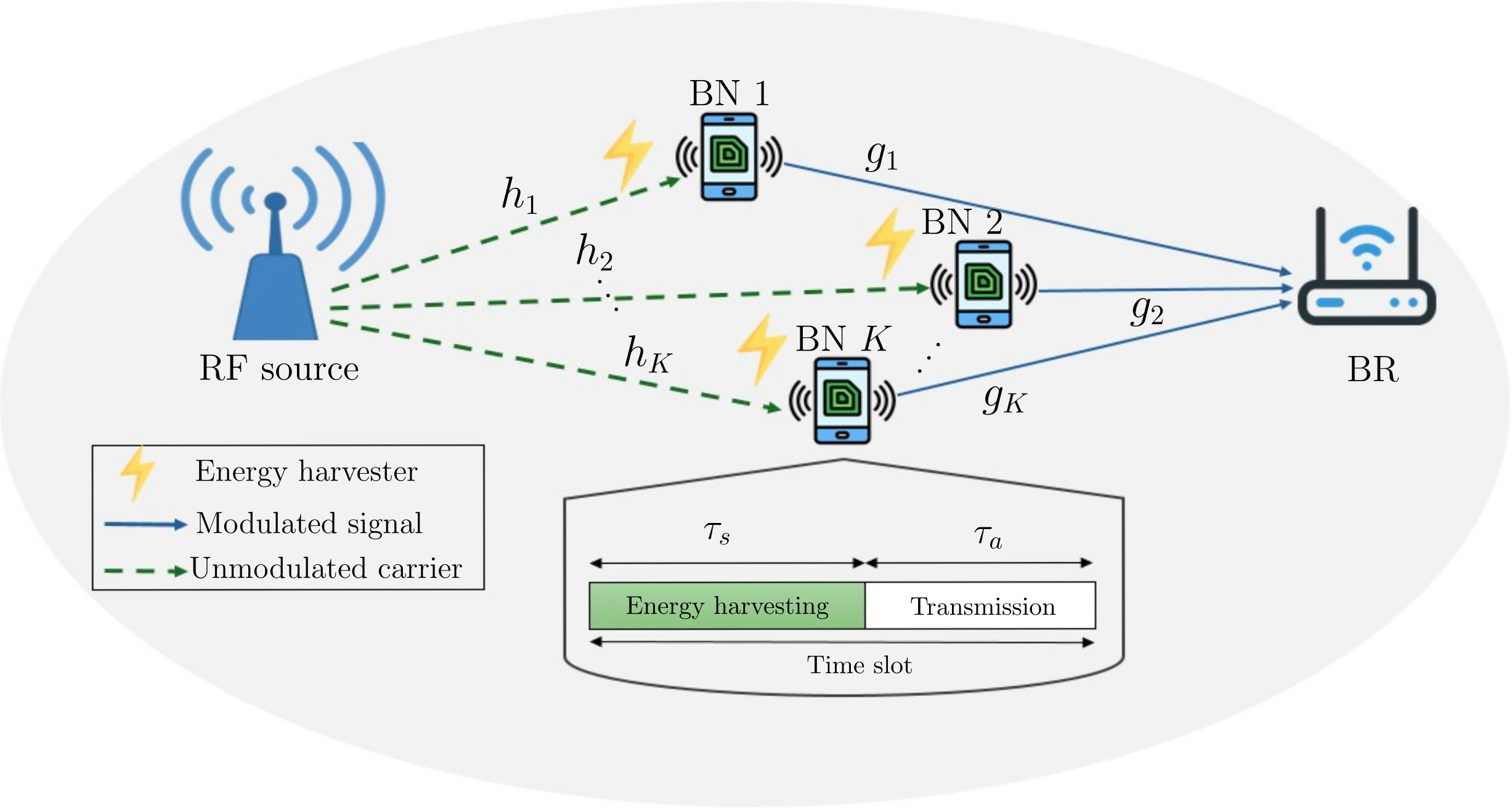}
    \caption{Uplink bistatic BackCom system with NOMA multiplexing}
    \label{fig:sys-model}
\end{figure}

We consider a NOMA-based bistatic BackCom network consisting of a dedicated RF energy source (e.g., an energy beacon or IoT gateway), a central backscatter receiver (BR) such as an access point, and a set of \( K \) batteryless backscatter nodes (BNs) \footnote{Note that while NOMA can theoretically support multiple BNs, practical challenges such as increased SIC complexity and the difficulty of achieving distinct received power levels in passive systems, make it challenging to support more than two BNs reliably. However, for a more comprehensive and complete analysis, this paper considers a general multi-BN case.}, as illustrated in Fig~\ref{fig:sys-model}. This architecture captures practical IoT scenarios such as warehouse monitoring or smart building sensing, where ultra-low-power tags operate
without batteries. 

The RF source continuously broadcasts an unmodulated carrier \footnote{Note that the RF source may transmit a known modulated signal $s(t)$. In this case, since it serves as a source of energy, $s(t)$ can be a predefined pattern that BR knows. Hence, it can be subtracted from the received signal. The backscattered signal then becomes a multiplication of $s(t)$ and the BN's signal. A common modeling approach approximates the product as a complex Gaussian random variable, and use the Shannon capacity formula to approximate the maximum achievable throughput \cite{shannon_apprx2}. The resulting throughput expression remains the same.} that serves two functions: (i) it provides energy for the BNs to harvest and (ii) it acts as a reference for backscatter modulation. Each BN encodes information by adjusting its load impedance to modulate the incident carrier and reflect it to the BR. A short beacon pilot at the start of each slot
provides time and frequency synchronization so that all BNs can reflect simultaneously, enabling practical SIC.

Each slot, normalized to unit duration and shorter than the channel coherence time, consists of a \textit{sleep phase} of length $\tau_s \in [0,1]$, during which BNs harvest energy, and an \textit{active phase} of length $\tau_a=1-\tau_s$, during which BNs backscatter data. Each BN uses a small capacitor to store harvested energy for the current slot and inter-slot storage is negligible \footnote{This assumption reflects practical designs of batteryless devices and avoids the need for long-term energy storage or state tracking, which will be considered for future work.}. 

Let $h_k$ and $g_k$ denote the complex channel coefficients from the RF source to BN $k$ and from BN $k$ to the BR, respectively, incorporating both path loss and Rayleigh fading. Shadowing is omitted for clarity, as it only scales the gains without changing the optimization framework. The BR orders BNs by descending order of the channel gain $|g_k|$ to facilitate SIC. Training is performed by assigning each BN a short reflection slot to estimate $|g_k|$ \footnote{During the training phase, each BN is assigned a dedicated time slot to reflect the carrier signal using the same reflection coefficient while others remain silent. The BR estimates the reflected link gains and orders the BNs accordingly.}.

During the sleep phase, each BN $k$ harvests
\begin{equation}
    E_k^{sleep}= \eta_k P_s |h_k|^2 \tau_s, \quad \forall{k}
\end{equation}
where $P_s$ is the RF-source power and $\eta_k \in [0,1]$ is the harvesting efficiency, which typically lies in the range $[0.4-0.7]$. During the active phase, each BN $k$ splits the incident RF power: a portion is used for backscatter communication, while the other is stored in the capacitor. The harvested energy is
\begin{equation}
    E_k^{active}= \eta_k (1-\beta_k) P_s |h_k|^2 \tau_a, \quad \forall{k} 
\end{equation}
where $\beta_k \in [0,1]$ is the reflection coefficient.

All BNs simultaneously reflect their signals using power-domain NOMA \footnote{NOMA requires simultaneous transmission. To preserve this in our model, a shared active phase is assumed, where BNs synchronize via a short beacon pilot broadcast at the start of each slot, providing the timing reference needed for SIC at the BR.}. The BR receives the superimposed signal expressed as
\begin{equation}
    y(t)= \sum_{k=1}^K  \sqrt{\beta_k P_s } \ h_k \ g_k \ x_k(t) + b(t)
\end{equation}
where $x(t)$ is the unit-power symbol of BN $k$, and $b(t)\sim \mathcal{CN}(0,\sigma^2)$ is the additive white Gaussian noise (AWGN). The BR applies SIC from the strongest channel gain $|g_k|$ to the weakest. Hence, the achievable rate of BN $k$ is 
\begin{equation}
    R_k = \tau_a \log_2 \left( 1+ \frac{\beta_k P_s \gamma_k}{\sum_{j>k} \beta_j P_s \gamma_j +1} \right), \quad \forall k,j 
\end{equation}
where $\gamma_k=|h_k|^2 |g_k|^2/\sigma^2$ and $\sum_{j>k}$ is the cumulative interference from BNs whose signals remain undecoded. The sum throughput is expressed as
\begin{equation}
    R_{sum}=\tau_a \log_2\left(1+ \sum_{k=1}^K \beta_k P_s \gamma_k\right)
\end{equation}

The total system energy consumption includes both the transmit energy of the RF source and the circuit energy of the RF source and BR. Hence, the total energy consumption per slot is
\begin{equation}
    E_{total}=\tau_s \left( \frac{P_s}{\xi} + P_{sc} \right) + \tau_a \left( \frac{P_s}{\xi} + P_{sc} + P_{rc} \right) 
\end{equation}
where $\xi \in [0,1]$ is the RF-source power amplifier efficiency, and $P_{sc}$ and $P_{rc}$ are the constant circuit powers of the RF source and BR, respectively. The BN circuit power is not included since it is supplied entirely by the harvested energy, which is already accounted for in the RF-source power consumption.

\section{Energy-efficient resource allocation}

\subsection{Problem formulation}

Maximizing EE is a key design challenge for energy-constrained backscatter IoT networks. We define EE as the ratio between the sum throughput and the system total energy consumption, measured in bits per Joule 
\begin{equation}
    EE=\frac{R_{sum}}{E_{total}}
\end{equation}

We jointly optimize the RF source transmit power $P_s$, the time allocation between sleep and actives phases ($\tau_a,\tau_s$),  and the reflection coefficient vector $\boldsymbol{\beta}$. 

Let $R_{sum} = \tau_a \cdot \tilde{R}_{sum}$ and $E_{total} = \tau_a \cdot \tilde{E}_{total}$. The EE reduces to  $\tilde{R}_{sum}/\tilde{E}_{total}$ and the optimization problem is formulated as
\begin{alignat*}{3}
    \mathbf{(EE0)} &\max_{P_s, \tau_a, \tau_s, \boldsymbol{\beta}} \quad && \frac{\tilde{R}_{sum}}{\tilde{E}_{total}} \\
    &\text{s.t.} \quad && \text{C1: } 0 \leq P_s \leq P_{max} \nonumber \\
    &&& \text{C2: } \tau_s \geq 0,\quad \tau_a > 0 \nonumber \\
    &&& \text{C3: } \tau_s + \tau_a = 1 \nonumber \\
    &&& \text{C4: } 0 \leq \beta_k \leq 1, \quad \forall k \nonumber \\
    &&& \text{C5: } P_{tc,k} \tau_a \leq E^{sleep}_k + E^{active}_k, \quad \forall k \nonumber
\end{alignat*}
where C1 limits the RF-source transmit power to a budget $P_{\max}$, and C5 ensures that the circuit power demand of BN $k$, given by $P_{tc,k}$, is met by the harvested energy.   

The problem is non-convex due to the fractional form of the objective function and the coupled variables. To solve this, we apply fractional programming techniques to transform the problem into a more tractable form, and then use an AO approach to solve it. 
\subsection{Dinkelbach-based AO approach}
The EE maximization $\mathbf{(EE0)}$ is a fractional programming problem. To solve it efficiently, we use Dinklebach's method which reformulates the ratio into a parameterized subtractive form. Specifically, the reformulated problem is expressed as
\begin{alignat*}{3}
    \mathbf{(EE1)} &\max_{P_s, \tau_a, \tau_s, \boldsymbol{\beta}} \quad && \tilde{R}_{sum} - \alpha \, \tilde{E}_{total} \\
    &\text{s.t.} \quad && \text{C1 - C5} \nonumber
\end{alignat*}
where $\alpha$ is a parameter that captures the tradeoff between the sum throughput and the total energy consumption,  and is iteratively updated by Dinkelbach’s algorithm until convergence.

We then, apply an AO approach, where each variable is optimized
sequentially while fixing the others. This allows to break the problem into simpler and tractable subproblems. We first solve $\mathbf{(EE1)}$ w.r.t  the reflection coefficient vector $\boldsymbol{\beta}$, assuming fixed values of $P_s$, $\tau_s$ and $\tau_a$.

\subsubsection{Reflection coefficients optimization}

It is easy to prove that the objective function in $\mathbf{(EE1)}$ is monotonically increasing with each reflection coefficient $\beta_k$. The feasible range is determined jointly by C4,  which imposes a simple box constraint, and C5. After simple mathematical manipulation of C5, we obtain the feasible upper bound as
\begin{equation}
    \beta_k \leq \frac{1}{\tau_a} - \frac{P_{tc,k}}{\eta_k P_s |h_k|^2}, \quad \forall k
\end{equation}

Combining constraints C4 and C5, the optimal reflection coefficient for BN $k$ is 
\begin{equation}
    \beta_k^*= \min \left(1, \max \left(\frac{1}{\tau_a} - \frac{P_{tc,k}}{\eta_k P_s |h_k|^2},0\right) \right), \quad \forall k
    \label{beta_opt}
\end{equation}

\subsubsection{Joint time allocation and power optimization}

Substituting $\beta^*_k$ in $\mathbf{(EE1)}$, the constraint C5 is satisfied, and the problem reduces to
\begin{alignat*}{3}
    \mathbf{(EE2)} &\max_{P_s, \tau_a, \tau_s} \quad && \tilde{R}_{sum} - \alpha \, \tilde{E}_{total} \\
    &\text{s.t.} \quad && \text{C1 - C3} \nonumber \\
    &&& \text{C6: } 0 \leq \frac{1}{\tau_a} - \frac{P_{tc,k}}{\eta_k P_s |h_k|^2} \leq 1, \ \ \ \forall k \nonumber 
\end{alignat*}

The subproblem is still coupled and non-convex. To address this, we introduce auxiliary variables to simplify the formulation. We denote
\begin{equation}
    \kappa= \left(1+ \frac{\tau_s}{1-\tau_s}\right) = \frac{1}{\tau_a}, \ \ \tilde{P_s}= \kappa P_s \label{aux_variables}
\end{equation}

Substituting \eqref{aux_variables} in $\mathbf{(EE2)}$, the constraint \text{C3} is satisfied and the subproblem is reformulated as
\begin{alignat*}{3}
    \mathbf{(EE3)} &\max_{\tilde{P_s},\, \kappa} \quad && \log_2 \left(\mu + \sum_{k=1}^K \gamma_k \tilde{P_s} \right) - \alpha \left( \frac{\tilde{P_s}}{\xi} + \kappa P_{sc}  + P_{rc} \right) \\
    &\text{s.t.} \quad && \text{C7: } 0 \leq \tilde{P_s} \leq \kappa P_{max}  \nonumber \\
    &&& \text{C8: } \kappa \geq 1 \nonumber \\
    &&& \text{C9: } 0 \leq \kappa \left( 1 - \frac{P_{tc,k}}{\eta_k \tilde{P_s} |h_k|^2} \right) \leq 1, \ \ \ \forall k \nonumber
\end{alignat*}
where $\mu= \left(1 - \sum_{k=1}^K \frac{P_{tc,k }\gamma_k}{\eta_k |h_k|^2} \right)$. For a fixed $\tilde{P_s}$, we observe that the objective function is monotonically decreasing with $\kappa$, and the optimal value lies at the lower bound of its feasible interval, which is determined by the power constraints
\begin{equation}
    \max \left(1, \frac{\tilde{P_s}}{P_{\max}} \right) \leq \kappa \leq \frac{1}{\left( 1- \frac{P_{tc,k}}{\eta_k \tilde{P_s} |h_k|^2} \right)} 
\end{equation}

Hence, the subproblem $\mathbf{(EE3)}$ can be equivalently divided into two subproblems corresponding to two operating modes depending on $\tilde{P}_s/P_{\max}$.

\vspace{0.1in}

\textit{i) Harvest-on-Transmit (HoT) mode}
\vspace{0.05in}

When $(\tilde{P_s}/P_{max}) \leq 1$, then $\kappa^* = 1$, which implies $\tau_s^*=0$ and $\tau_a^*=1$. The BNs operate without a dedicated sleep phase as the energy harvested during the active phase is sufficient to meet their circuit demands. This corresponds to a continuous operation mode without idle periods, which reduces latency. In this case, the subproblem simplifies to
\begin{alignat*}{3}
    \mathbf{(EE3i)} &\max_{\tilde{P_s}} \quad && \log_2 \left(\mu + \sum_{k=1}^K \gamma_k \tilde{P_s} \right) - \alpha \left( \frac{\tilde{P_s}}{\xi} + P_{sc} + P_{rc} \right) \\
    &\text{s.t.} \quad && \text{C10: } 0 \leq \tilde{P_s} \leq P_{max} \nonumber \\
    &&& \text{C11: } \tilde{P_s} \geq \frac{P_{tc,k}}{\eta_k |h_k|^2}, \ \ \ \forall k \nonumber
\end{alignat*}

The objective function is concave in $\tilde{P_s}$, and the optimal solution can be obtained either at the critical point $\tilde{P_s}^{(0,\mathrm{i})}= \xi/(\alpha \ln 2) - \mu/{\sum_{k=1}^K \gamma_k}$ by setting the derivative to zero, or at the boundaries of the feasible set. Hence, the solution is obtained in closed-form as
\begin{equation}
    \tilde{P_s}^*= \min \left(P_{\max}, \max \left(\tilde{P_s}^{(\min,\mathrm{i})}, \tilde{P_s}^{(0,\mathrm{i})}\right) \right) \label{sol:2i}
\end{equation}
where $\tilde{P_s}^{(\min,\mathrm{i})} = \max_k \left(P_{tc,k}/(\eta_k |h_k|^2)\right)$ is the minimum boundary. This closed-form solution highlights that in the HoT mode, BNs can operate efficiently and with reduced latency, but requires sufficient power during the active phase.

\vspace{0.1in}

\textit{ii) Harvest-then-Transmit (HtT) mode}

\vspace{0.05in}

When $(\tilde{P_s}/P_{max}) \geq 1$, then $\kappa^* = \tilde{P_s}/P_{\max}$, implying that $\tau_s^* > 0$ and $0< \tau_a^* <1$. A portion of the time slot is allocated to sleep, during which BNs harvest additional energy to complement what is collected during the active phase. This introduces latency but ensures sufficient energy is harvested to support active transmission. 
Based on \eqref{aux_variables}, the RF source must operate at maximum transmit power, i.e., $P_s^* = P_{\max}$, to ensure sufficient energy harvesting. Substituting $\kappa^* = \tilde{P_s}/P_{\max}$ in $\mathbf{(EE3)}$ reduces the subproblem to
\begin{alignat*}{3}
   \mathbf{(EE3ii)}&\max_{\tilde{P_s}} \quad &&   \! \log_2 \!\left(\!\mu \!+\! \sum_{k=1}^K \gamma_k \tilde{P_s} \!\right)\! - \!\alpha \! \left(\! \frac{\tilde{P_s}}{\xi} \!+\! \frac{P_{sc}}{P_{\max}}\tilde{P_s} \!+\! P_{rc} \!\right) \\
    &\text{s.t.} \quad &&  \text{C12: } \tilde{P_s}^{(\min,\mathrm{ii})} \leq \tilde{P_s} \leq \tilde{P_{s}}^{(\max)} \nonumber
\end{alignat*}
where $\tilde{P_s}^{(\min,\mathrm{ii})} = \max \left(P_{\max}, \max_k \left(P_{tc,k}/(\eta_k |h_k|^2)\right) \right)$ and    $\tilde{P_{s}}^{(\max)}=  P_{\max} + \min_k \left(P_{tc,k}/(\eta_k |h_k|^2)\right)$ are the minimum and maximum boundaries, respectively. Similar to the HoT case, the objective function is concave in $\tilde{P}_s$, and the closed-form solution is given by
\begin{equation}
    \tilde{P_s}^*= \min \left(\tilde{P_s}^{(\max)} , \max \left(\tilde{P_s}^{(\min,\mathrm{ii})}, \tilde{P_s}^{(0,\mathrm{ii})}\right) \right) \label{sol:2ii}
\end{equation}
where $\tilde{P_s}^{(0,\mathrm{ii})}= 1/\left(\alpha \ln 2 (\frac{1}{\xi} + \frac{P_{sc}}{P_{\max}})\right) - \mu/{\sum_{k=1}^K \gamma_k}$ is the critical point. This closed-form solution highlights that
in the HtT mode, BNs can remain active by dedicating a sleep phase to harvest sufficient energy, but this comes at the expense of reduced active time and increased latency.

Cases i) and ii) correspond to two operating modes in energy-harvesting backscatter networks. The algorithm to solve the optimization problem is given in Algorithm 1. This dual-mode operation captures the fundamental tradeoff between throughput, latency, and energy availability in energy-harvesting IoT networks.

\subsection{Convergence and complexity analysis}
We analyze the convergence of the proposed Dinkelbach-based AO algorithm. For a fixed parameter $\alpha$, the problem is decomposed into subproblems with closed-form solutions, each leads to an improvement over the previous iteration. This ensures convergence of each subproblem individually. Additionally, Dinkelbach’s algorithm is known to achieve superlinear convergence \cite{dinkelbach-convergence}. As a result, the EE values form a monotonically non-decreasing sequence, and since the objective is upper bounded by the RF source power and the energy harvesting constraints, the algorithm is guaranteed to converge.

The computational complexity per Dinkelbach iteration is $\mathcal{O}(K)$, which is dominated by the evaluation of constraint C4 and the computation of the reflection coefficient for each BN. Assuming the convergence is reached within $L$ iterations, 
the overall complexity is $\mathcal{O}(L \cdot K)$. This linear scaling complexity ensures that the algorithm remains efficient and suitable for large-scale IoT deployments.

\begin{algorithm}[t]
\caption{Dinkelbach-based AO algorithm}
\begin{algorithmic}[1]
\State \textbf{Input:} convergence threshold $\varepsilon > 0$, maximum iterations $L_{\max}$, initialize $\alpha = 0$, $\ell = 0$.
\Repeat
    \State \textbf{Step 1}: solve subproblem (EE3i) 
    \State Compute $\tilde{P}_s^{(i)}$ using \eqref{sol:2i} and the corresponding objective value $f^{(i)}$
    
    \State \textbf{Step 2}: solve subproblem (EE3ii) 
    \State Compute $\tilde{P}_s^{(ii)}$ using \eqref{sol:2ii} and the corresponding objective value $f^{(ii)}$
    
    \State \textbf{Step 3 (Mode selection)}:
    \If{$f^{(i)} > f^{(ii)}$}
        \State $\tilde{P}_s^* = \tilde{P}_s^{(i)}$, \quad $\kappa^* = 1$ $\rightarrow$ (HoT mode)
    \Else
        \State $\tilde{P}_s^* = \tilde{P}_s^{(ii)}$, \quad $\kappa^* = \tilde{P}_s^* / P_{\max}$ $\rightarrow$ (HtT mode)
    \EndIf

    \State \textbf{Step 4 (Dinkelbach update)}:
    \State Compute: $f(\alpha) = \log_2 \left(\mu + \sum_{k=1}^K \gamma_k \tilde{P_s} \right) - \ \ \alpha \left( \frac{\tilde{P_s}}{\xi} + \kappa P_{sc}  + P_{rc} \right)$
    \If{$|f(\alpha)| < \varepsilon$ or $l \geq L_{\max}$}
        \State Stop and return $\tilde{P_s}^*$ and $\kappa^*$
    \Else
        \State Update: $\alpha \gets \frac{\log_2 \left(\mu + \sum_{k=1}^K \gamma_k \tilde{P_s}^* \right)}{ \frac{\tilde{P_s}^*}{\xi} + \kappa^* P_{sc}  + P_{rc} }$
        \State $\ell \gets \ell + 1$
    \EndIf
\Until{convergence}

\State \textbf{Output} $P_s^* = \tilde{P}_s^* / \kappa^*$, \quad $\tau_a^* = 1 / \kappa^*$, \quad $\tau_s^* = 1 - \tau_a^*$
\State Compute $\beta_k^*$ using \eqref{beta_opt}
\end{algorithmic}
\end{algorithm}

\section{Simulation results}
\addtolength{\topmargin}{0.05in} 
The wireless channels are modeled as quasi-static Rayleigh fading with distance-dependent path loss. Specifically, the channel coefficients are given by  \( h_k = \tilde{h}_k d_{0,k}^{-n} \) and \( g_k = \tilde{g}_k d_k^{-n} \), 
where \( \tilde{h}_k, \tilde{g}_k \sim \mathcal{CN}(0,1) \) represent independent small-scale fading coefficients, $n$ is the path loss exponent, and \( d_{0,k} \), \( d_k \) are the distances from the RF source and the BR to BN $k$, respectively. The distances are chosen to introduce channel variation among BNs, with the maximum RF source-BR distance set to $40$~m. The simulation parameters are summarized in Table I, unless stated otherwise. All results are averaged over $N=10^5$ independent channel realizations via Monte Carlo simulation.

\begin{table}[htbp]
\centering
\caption{Simulation Parameters}
\begin{tabular}{l c}
\hline
\textbf{Parameter} & \textbf{Value} \\ \hline
Noise power $\sigma^2$ & $-100$ dBm \\
Harvesting efficiency $\eta_k$ & $0.6$ \\
Path loss exponent $n$ & $3$ 
\\
Power amplifier efficiency $\xi$ & $0.9$ \\
RF source circuit power $P_{sc}$ & $20$~Bm ($100$ mW) \\
BR circuit power $P_{rc}$ & $10$~dBm ($10$ mW) \\
BN circuit power $P_{tc,k}$ & $0$~dBm ($1$ mW) \\
\hline
\end{tabular}
\end{table}

\begin{figure}[htbp]
    \centering
    \subfloat[]{%
        \includegraphics[width=0.47\linewidth]{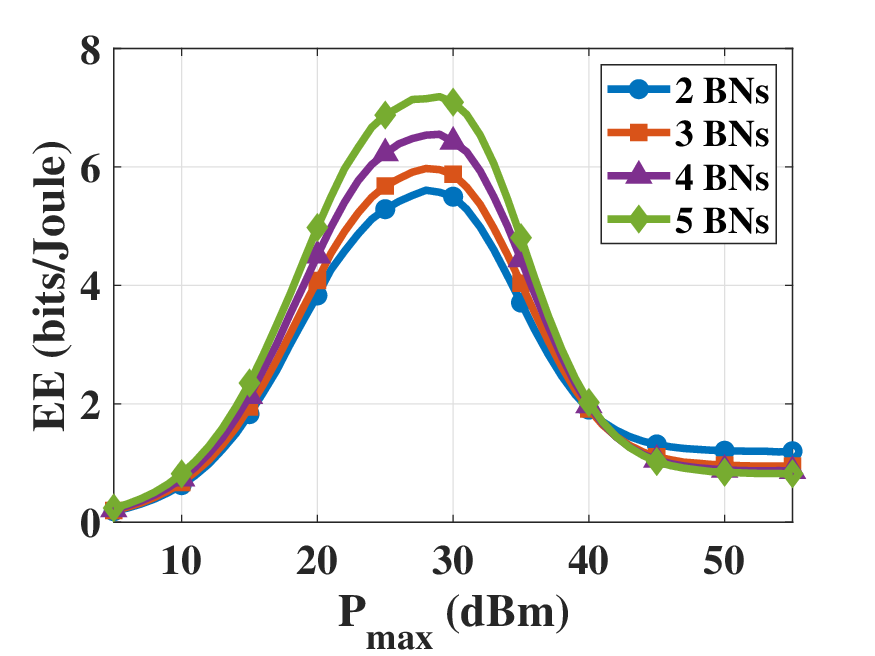}
        \label{fig:EE_users}
    }
    \hfill
    \subfloat[]{%
        \includegraphics[width=0.47\linewidth]{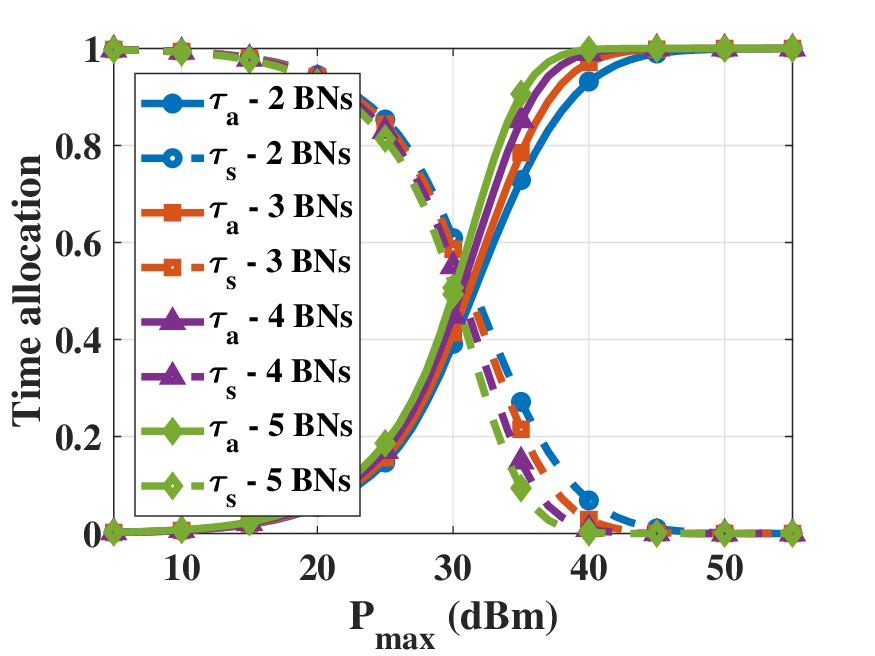}
        \label{fig:time_users}
    } \\
    \subfloat[]{%
        \includegraphics[width=0.47\linewidth]{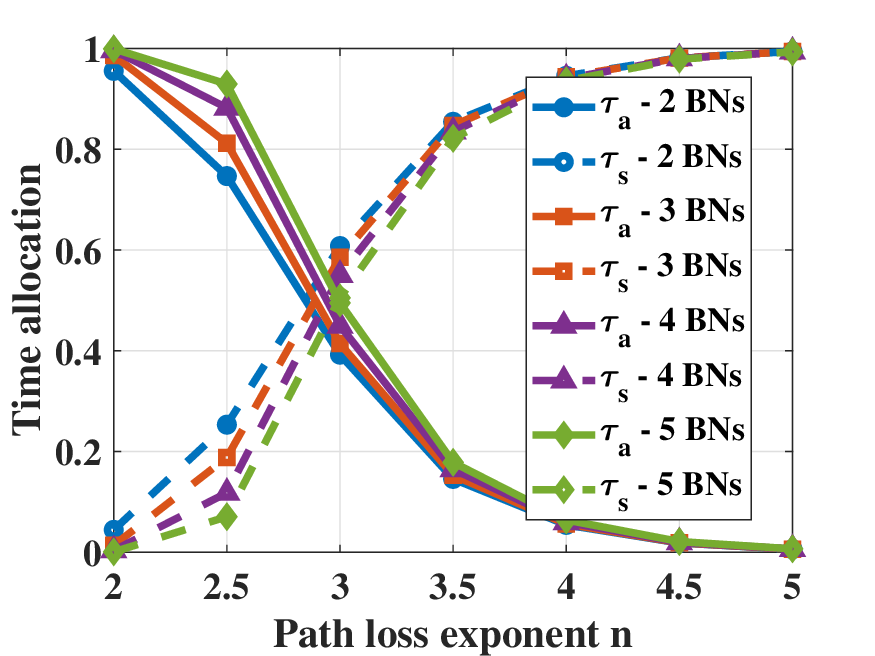}
        \label{fig:pathloss}
    }
    \hfill
    \subfloat[]{%
        \includegraphics[width=0.47\linewidth]{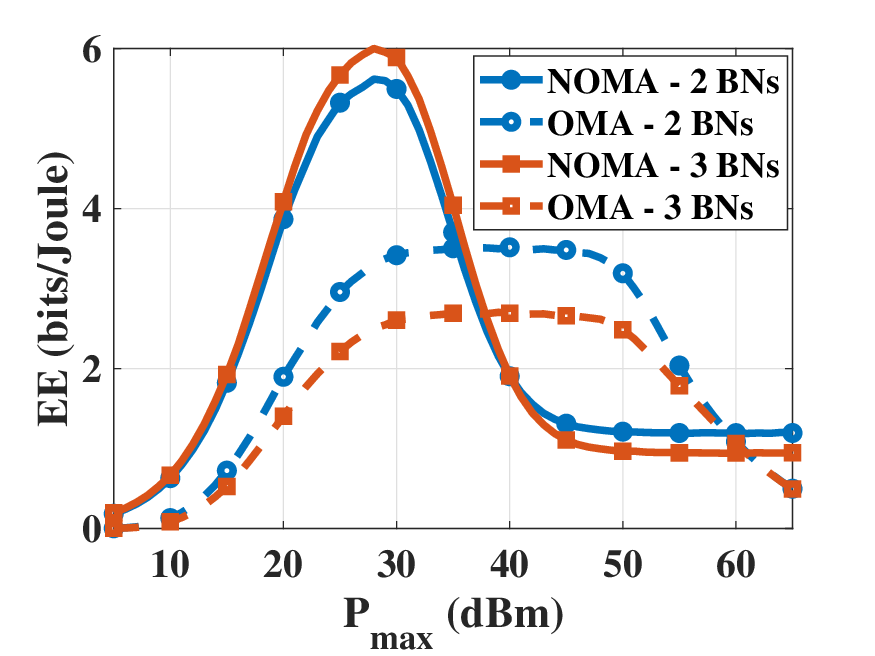}
        \label{fig:EE_OMA}
    }
    
    \caption{Performance evaluation: (a) EE vs. RF source power for different numbers of BNs; (b) Time allocation vs. RF source power for different numbers of BNs; (c) Time allocation vs. path loss exponent for different numbers of BNs; (d) EE comparison between NOMA and OMA schemes.}
    \label{fig:rf_budget_comparison}
\end{figure}

Fig.~\ref{fig:EE_users} shows the impact of the RF source power budget on EE for different numbers of BNs. Initially, increasing the RF source power leads to higher EE irrespective from the number of BNs.  However, beyond a certain threshold, i.e., $30$~dBm, EE starts to decrease. This is because additional transmit power yields only marginal throughput gains while significantly increasing energy consumption. Systems with more BNs achieve higher EE peak values, since the aggregate backscatter gain enhances the throughput without increasing the system circuit power as it is independent of the number of BNs. These results highlight the scalability of the proposed framework.

\begin{figure*}[htbp]
    \centering
    \subfloat[]{%
        \includegraphics[width=0.31\linewidth]{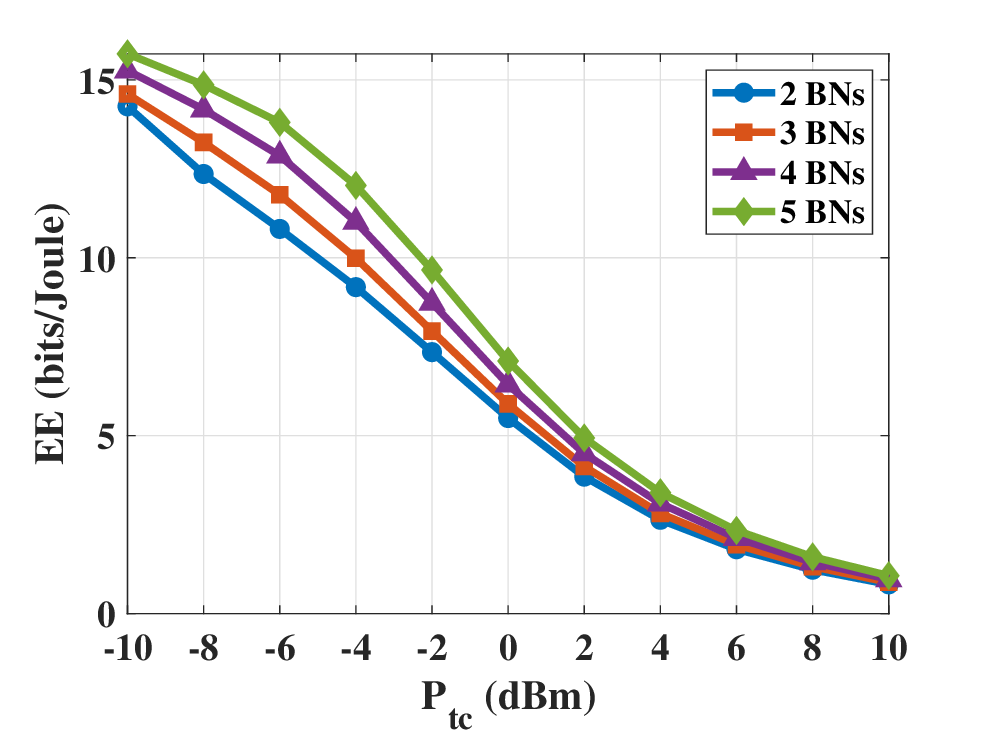}
        \label{fig:EE_Ptc}
    }
    \hfill
    \subfloat[]{%
        \includegraphics[width=0.31\linewidth]{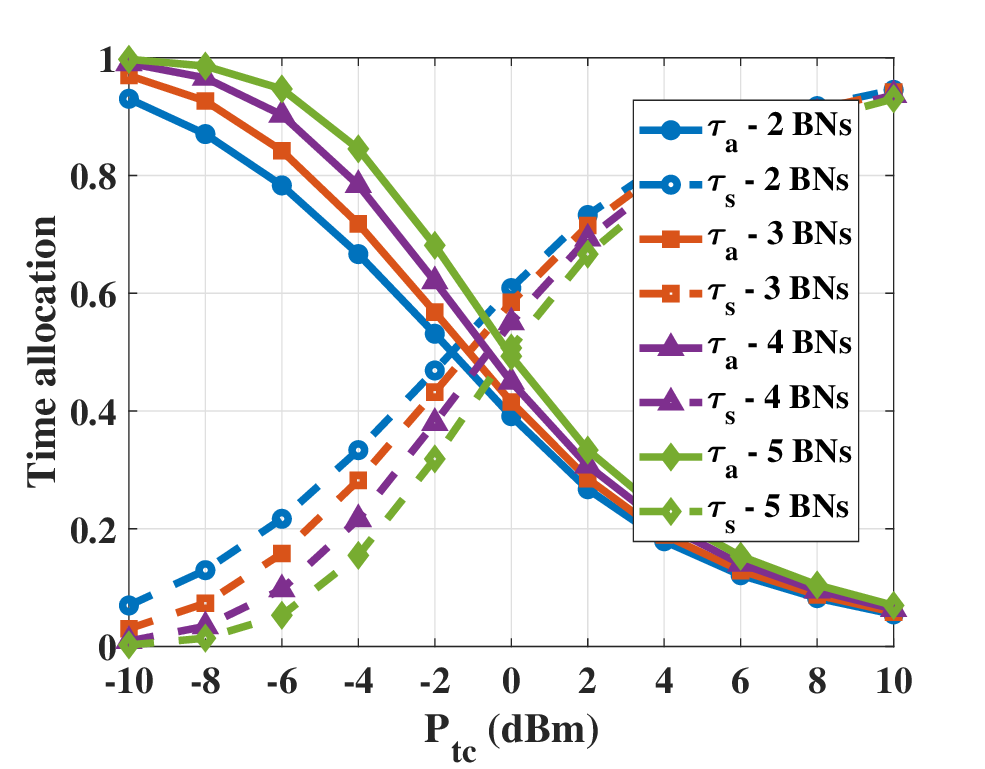}
        \label{fig:time_users_Ptc}
    }
    \hfill
    \subfloat[]{%
        \includegraphics[width=0.31\linewidth]{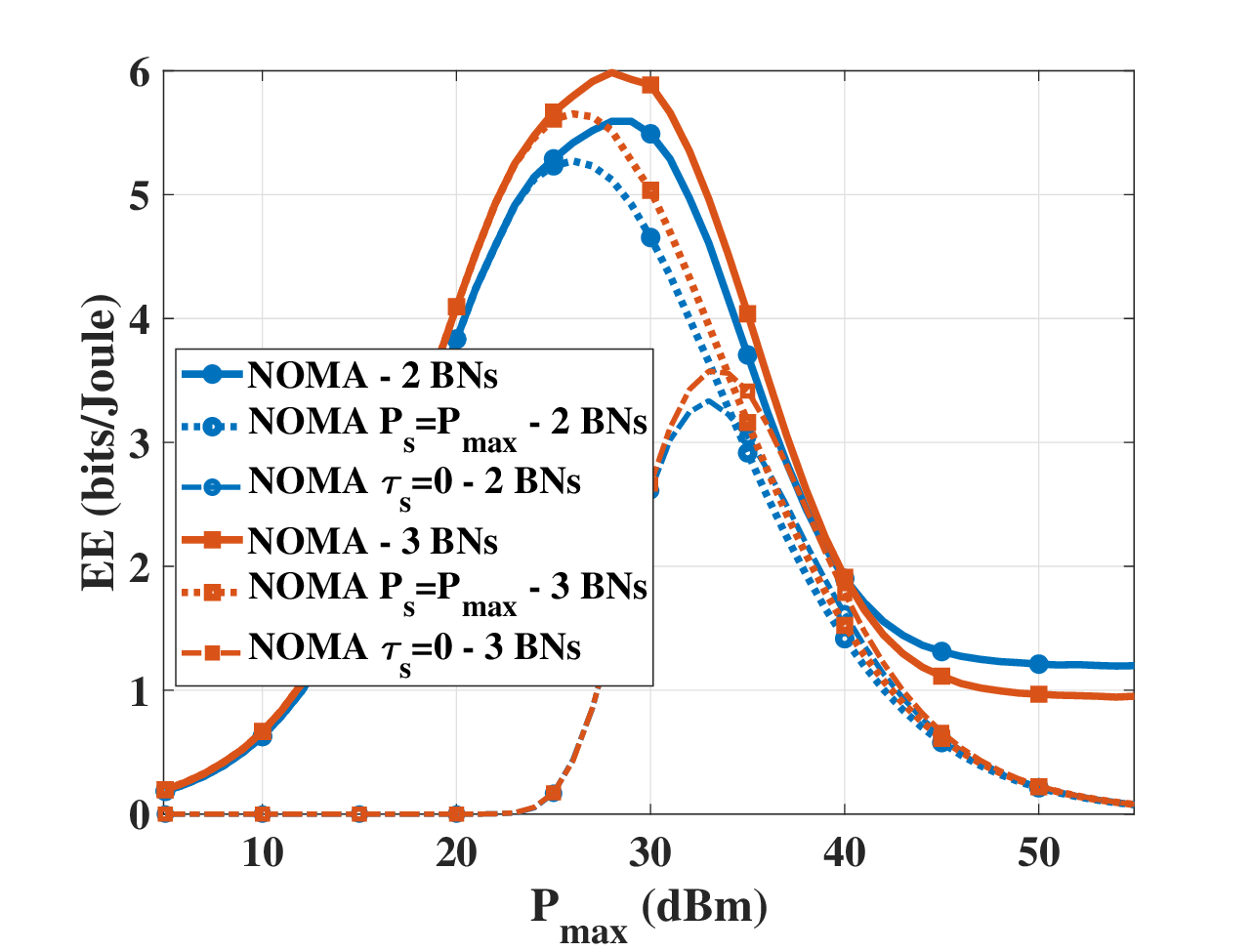}
        \label{fig:EE_different_methods}
    }
    
    \caption{Performance evaluation: (a) EE vs. BN circuit power for different numbers of BNs; (b) Time allocation vs. BN circuit power for different numbers of BNs; (c) EE comparison with different NOMA baselines.}
    \label{fig:bn_pathloss_results}
\end{figure*}

The corresponding time allocation is shown in Fig.~\ref{fig:time_users}. At low RF power levels ($\approx5-15$~dBm), BNs remain almost entirely in sleep phase $( \tau_s \approx 1$), as the harvested energy during the active phase is insufficient to sustain its active transmission. As the RF power increases to moderate levels ($\approx 15-40$~dBm), the sleep duration gradually decreases, and the active duration of BNs is extended. This behavior corresponds to the HtT mode, and the EE peak occurs when the sleep and active phases are allocated equally. Systems with more BNs require longer active durations because of the increased aggregate backscatter gain, which increases the contribution of the active phase to the sum throughput without additional energy consumption. At high RF power (beyond $\approx 40$~dBm), the system transitions into the HoT mode, where $\tau_a \approx1$ and the energy harvested during the active phase alone becomes sufficient for transmission.

Fig.~\ref{fig:pathloss} illustrates the impact of the path loss exponent on time allocation. For small values of $n$, corresponding to favorable propagation conditions, BNs remain active for most of the slot duration. As $n$ increases, the channels experience greater attenuation, reducing the energy harvested during the active phase. Consequently, BNs require longer dedicated sleep phases to accumulate sufficient energy, forcing them into the HtT mode under harsh propagation conditions. This highlights the sensitivity of backscatter operation to propagation conditions.

Fig.~\ref{fig:EE_OMA} compares the EE of NOMA and OMA schemes. NOMA consistently achieves higher EE at low and moderate power levels, since simultaneous transmissions during the active phase improve the sum throughput. However, at high power levels, OMA outperforms NOMA, but NOMA sill achieves the highest EE peak around $P_{\max}=30$~dBm, with up to $127\%$ gain over OMA, highlighting its advantage in energy-constrained scenarios.

The impact of BN circuit power consumption is shown in Figs.~\ref{fig:EE_Ptc} and \ref{fig:time_users_Ptc} for $P_{\max}=30$~dBm. When the circuit power requirement is low, BNs can remain active for longer durations. However, as the circuit power demand increases, the active duration shortens and the sleep duration is extended, allowing BNs to harvest additional energy to meet their higher circuit demands. As a result, the EE decreases as more energy is consumed by the circuitry without proportional gain in the throughput.

Finally, Fig.~\ref{fig:EE_different_methods} compares the proposed scheme with two baselines: \textit{(i) fixed RF power} with $P_s = P_{\max}$, in dashed lines, and \textit{(ii) no sleep phase} with $\tau_s = 0$, in dash-dotted lines. At low RF power levels, the fixed-power baseline perform similarly to the proposed scheme, since operating at full power is necessary to harvest sufficient energy. However, as the power increases, this strategy becomes inefficient since the excessive power increases energy consumption without proportional gain in the throughput. The no-sleep baseline performs worst at low and moderate power levels, where the absence of a sleep phase limits the energy available for BNs transmission. The performance improves at high power levels, but remains below the proposed scheme. Overall, the proposed scheme achieves up to $68\%$ higher EE peak compared to the no-sleep baseline and $8\%$ over the fixed-power baseline.


\section{Conclusions}
This paper presented an energy-efficient resource allocation framework for uplink NOMA-enabled backscatter communication networks. The joint optimization of RF source transmit power, backscatter reflection coefficients, and sleep/active time allocation was formulated as a non-convex fractional program. To solve it, we developed a low-complexity Dinkelbach-based AO algorithm with guaranteed convergence. The analysis revealed two fundamental operating modes: harvest-on-transmit (HoT), where BNs operate with low latency when RF power is sufficiently high, circuit demands are modest, and channels are favorable; and harvest-then-transmit (HtT), where longer dedicated sleep phase is necessary under limited RF power budget, higher circuit demands or unfavorable channels. Simulation results showed improvement in EE compared to fixed-power and no-sleep baselines, with NOMA outperforming OMA in low- and moderate-power regimes. These findings highlight the benefits of adaptive mode selection, supported by power control and sleep-active scheduling, for the scalability of battery-free IoT deployments, and point to future extensions on robustness under imperfect channel state information, non-ideal energy harvesters, and multi-cell scenarios for 6G backscatter networks.

\bibliographystyle{IEEEtran}
\bibliography{IEEEabrv.bib, biblio.bib}
\end{document}